\documentclass[12pt]{article}
\usepackage{amsmath}
\usepackage[dvips]{graphicx}
\input amssym

\begin{document}
	
	\def\inn{\,\rfloor\,}                  \def\bt{{\bar\tau}}
	\def\br{{\bar\rho}}
	\def\prg#1{\medskip\noindent{\bf #1}}  \def\ra{\rightarrow}
	\def\lra{\leftrightarrow}              \def\Ra{\Rightarrow}
	\def\nin{\noindent}                    \def\pd{\partial}
	\def\dis{\displaystyle}                \def\inn{\,\rfloor\,}
	\def\grl{{GR$_\Lambda$}}               \def\Lra{{\Leftrightarrow}}
	\def\cs{{\scriptstyle\rm CS}}          \def\ads3{{\rm AdS$_3$}}
	\def\Leff{\hbox{$\mit\L_{\hspace{.6pt}\rm eff}\,$}}
	\def\bull{\raise.25ex\hbox{\vrule height.8ex width.8ex}}
	\def\ric{{(Ric)}}                      \def\tric{{(\widetilde{Ric})}}
	\def\tmgl{\hbox{TMG$_\Lambda$}}
	\def\Lie{{\cal L}\hspace{-.7em}\raise.25ex\hbox{--}\hspace{.2em}}
	\def\sS{\hspace{2pt}S\hspace{-0.83em}\diagup}   \def\hd{{^\star}}
	\def\dis{\displaystyle}     \def\ul#1{\underline{#1}}
	\def\pgt{{\scriptstyle\rm PGT}}
	\def\ul{{\underline}}
	
	\def\hook{\hbox{\vrule height0pt width4pt depth0.3pt
			\vrule height7pt width0.3pt depth0.3pt
			\vrule height0pt width2pt depth0pt}\hspace{0.8pt}}
	\def\semidirect{\;{\rlap{$\supset$}\times}\;}
	\def\first{\rm (1ST)}       \def\second{\hspace{-1cm}\rm (2ND)}
	\def\bm#1{\hbox{{\boldmath $#1$}}}      \def\un#1{\underline{#1}}
	\def\nb#1{\marginpar{{\large\bf #1}}}
	\def\diag{{\rm diag}\,}
	
	\def\G{\Gamma}        \def\S{\Sigma}        \def\L{{\mit\Lambda}}
	\def\D{\Delta}        \def\Th{\Theta}
	\def\a{\alpha}        \def\b{\beta}         \def\g{\gamma}
	\def\d{\delta}        \def\m{\mu}           \def\n{\nu}
	\def\th{\theta}       \def\k{\kappa}        \def\l{\lambda}
	\def\vphi{\varphi}    \def\ve{\varepsilon}  \def\p{\pi}
	\def\r{\rho}          \def\Om{\Omega}       \def\om{\omega}
	\def\s{\sigma}        \def\t{\tau}          \def\eps{\epsilon}
	\def\nab{\nabla}      \def\btz{{\rm BTZ}}   \def\heps{\hat\eps}
	\def\bu{{\bar u}}     \def\bv{{\bar v}}     \def\bs{{\bar s}}
	\def\te{{\tilde e}}   \def\tk{{\tilde k}}
	\def\bare{{\bar e}}   \def\bark{{\bar k}}
	\def\barom{{\bar\omega}}  \def\barg{{\bar g}}
	\def\tphi{{\tilde\vphi}}  \def\tt{{\tilde t}}
	\def\te{{\tilde e}}
	
	\def\tG{{\tilde G}}   \def\cF{{\cal F}}    \def\bH{{\bar H}}
	\def\cL{{\cal L}}     \def\cM{{\cal M }}   \def\cE{{\cal E}}
	\def\cH{{\cal H}}     \def\hcH{\hat{\cH}}
	\def\cK{{\cal K}}     \def\hcK{\hat{\cK}}  \def\cT{{\cal T}}
	\def\cO{{\cal O}}     \def\hcO{\hat{\cal O}} \def\cV{{\cal V}}
	\def\tom{{\tilde\omega}}  \def\cE{{\cal E}}     \def\tR{{\tilde R}}
	\def\cR{{\cal R}}    \def\hR{{\hat R}{}}   \def\hL{{\hat\L}}
	\def\tb{{\tilde b}}  \def\tA{{\tilde A}}   \def\tv{{\tilde v}}
	\def\tT{{\tilde T}}  \def\tR{{\tilde R}}   \def\tcL{{\tilde\cL}}
	\def\hy{{\hat y}}    \def\tcO{{\tilde\cO}} \def\hom{{\hat\om}}
	\def\he{{\hat e}}
	\def\nn{\nonumber}                    \def\vsm{\vspace{-10pt}}
	\def\be{\begin{equation}}             \def\ee{\end{equation}}
	\def\ba#1{\begin{array}{#1}}          \def\ea{\end{array}}
	\def\bea{\begin{eqnarray} }           \def\eea{\end{eqnarray} }
	\def\beann{\begin{eqnarray*} }        \def\eeann{\end{eqnarray*} }
	\def\beal{\begin{eqalign}}            \def\eeal{\end{eqalign}}
	\def\lab#1{\label{eq:#1}}             \def\eq#1{(\ref{eq:#1})}
	\def\bsubeq{\begin{subequations}}     \def\esubeq{\end{subequations}}
	\def\bitem{\begin{itemize}}           \def\eitem{\end{itemize}}
	\renewcommand{\theequation}{\thesection.\arabic{equation}}

\title{Near-horizon geometry with torsion}
\author{ B. Cvetkovi\'c and D. Simi\'c\footnote{
        Email addresses: {cbranislav@ipb.ac.rs, dsimic@ipb.ac.rs}}\\
Institute of Physics, University of Belgrade \\
                      Pregrevica 118, 11080 Belgrade, Serbia}

\maketitle

\begin{abstract}

We investigate near-horizon geometry of  the rotating Ba\~nados Teiteilboim Zanelli (BTZ) black hole with torsion. Our main motivation is to gain insight into the role of torsion in the near-horizon geometry, which is well understood in the Riemannian case. We  obtain that near-horizon geometry
represents a  generalization of  AdS self-dual orbifold with non-trivial torsion. We analyze its asymptotic structure and derive the corresponding algebra of asymptotic symmetries, which consists of  chiral Virasoro and centrally extended $u(1)$ Kac-Moody algebra.

\end{abstract}

\section{Introduction}

Black holes represent some of the most fascinating objects in our Universe which, since their discovery by Schwarzschild, do not stop to puzzle and inspire us. After Hawking's discovery  of black hole radiation we stumbled into quite a few problems. First of them is an information paradox. Namely, if black hole radiates thermally,
 after its evaporation the information about the matter it was constituted of is inevitably lost. The second one  is the problem of black hole micro-states which are responsible for the black hole entropy and crucial for the   information paradox resolving.

There are numerous approaches to the previously mentioned problems, in this article we are particularly interested in Kerr/CFT \cite{x1}. The basic idea of Kerr/CFT is that near horizon geometry encodes many important information about the full geometry itself if not them all. The only drawback is that near-horizon geometry is well-defined only for the extremal black holes. Nevertheless, keeping in mind the importance of the subject, gaining diverse insights is valuable.

It is worth noting that charges that generate asymptotic symmetry of near-horizon geometry are by construction soft. This is interesting because there are propositions that soft hairs are black hole micro-states \cite{x2,x3}. So, one might hope to gain a small insight into importance of torsion for this approach as well.

Although Kerr/CFT is, already, known for a decade there is no analogous analysis for the gravity with torsion. Our intention is to fill this gap and initiate investigation of near-horizon geometries with torsion. Before proceeding to the subject of our work it is instructive to give a short note on the role of torsion in gravity.

Einstein's general relativity (GR), prototype of all modern gravitational theories, postulates that connection is Christoffel or, equivalently, that there is no torsion. Although successful in explaining  many of the existing macroscopic phenomena it still
lacks the status of the fundamental theory on the microscopic level, due its non-renormalizability. One also needs to include dark matter and dark energy into GR  to accomplish the agreement with  observations.

Instead of including mysterious new form of matter one may turn to alternative theories
of gravity. If we allow the presence of torsion in gravitational theories we are exposed to a vast number of new possibilities. One of them is that dark matter represents one of the manifestations of non-trivial torsion, see \cite{x4,x5,x6}. In cosmology there are also considerations in which both dark matter and energy are replaced by torsion \cite{x7}. Additionally, there are proposals that torsion plays a crucial role in inflation. Namely in \cite{x8} a simple model of inflation is proposed which, among other issues, also solves the problem of cosmological singularity.

However,  there is an approach to gravity based on localization of Poincar\'e group in which basic dynamical variables are vielbein and spin connection. For the comprehensive overview of the subject see \cite{x9}. This formulation of gravity naturally incorporates presence of torsion and metric postulate is a consequence of an antisymmetry of spin connection. Also coupling of fermions to gravity is natural, which  makes this approach more appealing formulation than standard metric one.

The paper is organized as follows. In the next section we collect main results of Mielke-Baekler (MB) model: action, equations
of motion, BTZ black hole with torsion, canonical structure and generator of gauge symmetries. Third section is devoted to the main results of this paper. We construct near-horizon geometry of BTZ black hole with torsion which represents a new solution of MB model, self-dual orbifold with torsion and give its basic properties. After introducing suitable asymptotic conditions for vielbein and spin connection we derive the kinematical algebra of asymptotic symmetries. Using the results of the second section we construct the corresponding charges that generate this symmetry and derive commutation relations between  generators.

Our conventions are as follows. The Latin indices $(i,
j, k, ...)$ refer to the local Lorentz frame with the
signature of the metric $(+,-,-)$, Levi-Civita symbol
$\ve^{ijk}$ is normalized to $\ve^{012}=1$. We use Greek indices
$(\m,\n,\r, ...)$ for the coordinate frame. The orthonormal
triad (coframe 1-form) is denoted with $e^i$, while $\om^{ij}$ is the spin connection (1-form). The field strengths are torsion $T^i=de^i+\om^i{_m}\wedge e^m$
and the curvature $R^{ij}=d\om^{ij}+\om^i{_k}\wedge \om^{kj}$ (2-forms). The Lie dual of an
antisymmetric form $X^{ij}$ is $X_i:=-\ve_{ijk}X^{jk}/2$ and the exterior product of forms is implicit.

\section{Mielke-Baekler model}
\subsection{Mielke-Baekler model in a nutshell}
Basic dynamical variables in the first order formulation of gravity are vielbein $e^i_{\ \mu}$ and spin connection $\omega^{ij}_{\ \mu}$. Very common is to use differential forms which, often, simplify notation and calculation, so we introduce vielbein and spin connection 1-forms
\bea
e^i=e^i_{\ \mu}dx^\mu\,, & \omega^{ij}=\omega^{ij}{}_{\mu}dx^\mu
\eea
In three dimensions (3D) it is useful to pass on dual spin connection
$\omega_i=-\frac{1}{2}\varepsilon_{ijk}\omega^{jk}$, and in completely same manner for every other two-index Lorentz tensors.
In this notation torsion and curvature 2-forms are given by
$$
T^i=de^i+\varepsilon^i_{\ jk}\omega^je^k\,,\qquad
R^i=d\omega^i+\varepsilon^i_{\ jk}\omega^j\omega^k.
$$
MB {\it topological} model \cite{x10} of 3D gravity is described by the action
\bea
&&I=aI_1+\L I_2+\a_3I_3+\a_4I_4+I_M\,,\nn\\
&&I_1=2\int e^iR_i\,,\qquad I_2=-\frac{1}{3}\int\ve_{ijk}e^ie^je^k\,,\nn\\
&&I_3=\int \omega_id\omega^i+\frac{1}{3}\varepsilon_{ijk}\omega^i\omega^j\omega^k\,,\qquad I_4=\int e^iT_i\,,
\eea
where $I_M$ is the matter field contribution. The first term with $a=1/16\pi G$ is the Einstein Cartan action, the second term is
the cosmogical one, the third terms is Chern-Simons action for Lorentz connection, while the fourth term explicitly depends on torsion.
In the previous expressions and in what follows wedge product between forms is implicit.

We are particularly interested in the non-degenerate sector of the theory in which the following relation holds
$
a^2-\a_3\a_4\neq 0.
$
If the previous relation is true then equations of motion can be cast in the simple form \cite{x11}
\bsubeq
\bea
&&2T_i=p\ve_{ijk}e^je^k\,,\qquad p=\frac{\a_3\varLambda+\a_4a}{\a_3\a_4-a^2}\,,\lab{2.3a}\\
&&2R_i=q\ve_{ijk}e^je^k\,,\qquad q=-\dis\frac{\a_4^2+a\varLambda}{\a_3\a_4-a^2}\lab{2.3b}\,.
\eea
\esubeq
The vacuum configuration is characterized by constant
torsion and constant curvature. For  $p=0$ , or $q=0$ the vacuum geometry is Riemannian ($T^i=0$) or teleparallel
($R^i=0$).

From the torsion equation \eq{2.3a} it follows that contorsion one form is particulary simple, $K^i=\frac p2 e^i$, so that
connection reads
\be
\om^i=\tom^i+\frac p2e^i\,,
\ee
where $\tom^i$ is Levi-Chivita (Riemannian) connection.  The curvature equation of motion \eq{2.3b}
now implies that  Riemannian piece of curvature is
\bsubeq
\be
2\tR_i=\Leff \ve_{ijk}e^ie^j\,,\qquad \Leff=q-\frac{p^2}4\,,
\ee
where $\Leff$ is an effective cosmological constant. In what follows we shall restrict ourselves to
the AdS sector of the theory with negative effective cosmological constant
\be
\Leff:=-\frac1{\ell^2}\,,
\ee
\esubeq
where $\ell$ is an AdS radius.

\subsection{BTZ black hole with torsion}

MB model has an important and interesting, solution which is a generalization of BTZ black hole and possesses non-trivial torsion \cite{x12}.

The metric of this solution, parametrized by $m$ and $j$, is given by
\bsubeq
\be
ds^2=N^2dt^2-\frac{dr^2}{N^2}-r^2(N_\varphi dt+d\varphi)^2\,,
\ee
where
\bea
N^2=-8Gm+\frac{r^2}{\ell^2}+\frac{16G^2j^2}{r^2}\,, & N_\varphi=\dis\frac{4Gj}{r^2}\,.
\eea
\esubeq
Triad fields can be chosen in a simple diagonal form
\bea
e^0=N dt\,,\quad e^1=\dis\frac{dr}{N}\,, \quad e^2=r(N_\varphi dt+d\varphi)\,,
\eea
The spin connection of the solution is
\be
\omega^i=\tilde{\omega}^i+\frac{p}{2}e^i
\ee
where $\tilde{\omega}^i$ is Levi-Chivita  connection.

Conserved charges of the black hole are energy
\be
E=m\left(a+\frac{\a_3p}{2}\right)-\frac{\a_3}{\ell^2}j\,,
\ee
and angular momentum
\be
J=\left(a+\frac{\a_3p}{2}\right)j-\a_3m\,.
\ee
The entropy of the solution contains a contribution stemming from torsion.
For more details  see \cite{x13}.

\subsection{Canonical generator of gauge symmetries}

We shall first review some of the results concerning the canonical structure of the MB model, for details  \cite{x11}.

\prg{Primary and secondary constraints.} There are six first class primary  constraints in the theory
\bsubeq
\bea
\pi_i^{\ 0}\approx0\,, \qquad \Pi_i^{\ 0}\approx0\,,
\eea
where $\pi_i{^\mu}$ and $\Pi_i{^\mu}$ are momenta
conjugate to basic dynamical variables $e^i{_\mu}$ and $\om^i{_\mu}$.
Secondary first class constraints in the reduced phase space, obtained
from the original phase space after elimination of second class constraints, are given by
\bea
\cH_i=-\ve^{0\a\b}\left(aR_{i\a\b}+\a_4T_{i\a\b}+\varLambda\varepsilon_{ijk}e^j_{\ \a}e^k_{\ \b}\right)\,,\nn \\
\cK_i=-\ve^{0\a\b}\left(aT_{i\a\b}+\a_3R_{i\a\b}+\a_4\varepsilon_{ijk}e^k_{\ \a}e^k_{\ \b}\right)\,,
\eea
\esubeq

\prg{Canonical generator.} We shall now state the form of  the generator of gauge symmetry, which will be used in the next section for determination of the asymptotic symmetries. The general procedure for  the generator construction is given in  \cite{x14}. Generator of gauge symmetries $G$ consists of two parts
$$
G=-G_1-G_2\,.
$$
The first part generates diffeomorphisms and has  the following form
\bsubeq
\be
G_1=\dot{\xi}^\rho\left(e^i_{\ \rho}\pi_i^{\ 0}+\omega^i_{\ \r}\Pi_i^{\ 0} \right)+\xi^\rho\left(e^i{_\rho}\cH_i+\omega^i_{\ \rho}\cK_i+
\left(\pd_\rho e^i_{\ 0}\right)\pi_i ^{\ 0}+\left(\pd_\rho\omega^i_{\ 0}\right)\varPi_i^{\ 0}\right)\,,
\ee
while the second generates local Lorentz transformations and is given by
\be
G_2=\dot{\theta}^i\varPi_i^{\ 0}+\theta^i\left(\cK_i-\varepsilon_{ijk}\left(e^j{_0}\pi^{k0}+\omega^j_{\ 0}\varPi^{k0}\right)  \right)\,.
\ee
\esubeq

\section{AdS self-dual orbifold with torsion}
\setcounter{equation}{0}

\subsection{Near-horizon of the BTZ with torsion}
We study the near-horizon limit of the extremal BTZ black hole with torsion.
Extremal BTZ black hole with torsion is characterized by the following relation between the parameters
\be
j=\pm m\ell,
\ee
from which we deduce the value of the radius of the event horizon
\bsubeq
\be
r_0=2\ell\sqrt{Gm},
\ee
and angular velocity
\be
\Om=N_\varphi(r_0)=\frac{4Gj}{r_0^2}=\frac1\ell\,.
\ee
\esubeq
In order to arrive at the  near-horizon region, we need to change the variables in the following manner
\bea
\vphi \ra \vphi - \Om t/\eps\,, & r\ra r_0 +r\eps \,,
 & t \rightarrow t/\epsilon.
\eea
and after taking the limit $\eps \ra 0$,
we derive the near-horizon geometry of the BTZ with torsion, which is characterized by
\bea
N^2dt^2 \rightarrow \frac{4r^2}{\ell^2} dt^2\,,\qquad \frac{dr^2}{N^2} \rightarrow \frac{\ell^2}{4r^2} dr^2\,,\qquad
d\varphi+N_\varphi dt\rightarrow d\varphi -\frac{2r}{r_0\ell} dt\,,\nn
\eea
so we arrive at the metric of the near-horizon
\be
ds^2=\frac{4r_0r}{\ell}dtd\varphi-\frac{\ell^2}{4r^2}dr^2-r^2_0d\varphi^2\,.
\ee
The triad fields  can be easily derived from the metric
\bea
e^0=\frac{2r}{\ell}dt\,,\qquad e^1=\frac{\ell}{2r}dr\,,\qquad e^2=-r_0d\varphi+\frac{2r}{\ell}dt\,.
\eea
The Levi-Chivita connection is given by
\bsubeq
\bea
\tom^0=-\frac{e^0}{\ell}\,,\qquad \tom^1=-\frac{e^1}{\ell}\,,\qquad \tom^2=-\frac{2e^0}{\ell}+\frac{e^2}{\ell}\,,
\eea
and the Cartan spin connection reads
\be
\omega^i=\tilde{\omega^i}+\frac{p}{2}e^i.
\ee
\esubeq
In this way we constructed the new solution of the MB model which represents the generalization of AdS$_3$ self-dual orbifold \cite{x15}, with non-trivial torsion.

\subsection{Asymptotic conditions}

To further explore properties of the AdS orbifold with torsion we analyze asymptotic  structure of this solution.
First, we introduce asymptotic conditions of the metric, which are
\bsubeq
\be
g_{\m\n}\sim\left(
\ba{ccc}
\cO_{0}&\cO_{3}&\cO_{-1}\\
\cO_{3}&\dis
-\frac{\ell^2}{4r^2}+\cO_4&\cO_1\\
\cO_{-1}&\cO_1&\cO_0
\ea
\right)\,,
\ee
where $\mathcal{O}_n$ stands for therm that at infinity behaves as $r^{-n}$.
The asymptotic behavior of the triad is given by
\bea
e^i{_\m}\sim\left(
\ba{ccc}
\dis\frac{2r}{\ell}+\cO_{1}&\cO_4&\cO_0\\
\cO_2&\dis\frac\ell {2r}+\cO_3&\cO_0\\
\dis\frac{2r}{\ell}+\cO_1&\cO_4&\cO_0
\ea
\right)\,.
\eea
\esubeq
Asymptotic form of the Levi-Chivita connection reads
\bsubeq
\bea
\tom^i{_\m}\sim\left(
\ba{ccc}
-\dis\frac{2r}{\ell^2}+\cO_1&\cO_2&\cO_0\\
\cO_1&-\dis\frac{1}{2r}+\cO_2&\cO_0\\
-\dis\frac{2r}{\ell^2}+\cO_{1}&\cO_2&\cO_0
\ea
\right)\,.
\eea
By using $\omega^i=\tom^i+\dis\frac{p}{2}e^i$ we conclude that asymptotic form of Cartan connection is
\bea
\omega^i{_\m}\sim\left(
\ba{ccc}
-\dis\frac{2r}{\ell^2}+\frac{pr}{\ell}+\cO_1&\cO_2&\cO_0\\
\cO_1&-\dis\frac{1}{2r}+\frac{p\ell}{4r}+\cO_2&\cO_0\\
-\dis\frac{2r}{\ell^2}+\frac{pr}{\ell}+\cO_{1}&\cO_2&\cO_0
\ea
\right)
\eea
\esubeq
From the adopted asymptotic behavior of the fields we can derive kinematical algebra of asymptotic symmetries.
The most simple way of deriving sub-algebra of the algebra of diffeomorphisms under which the adopted asymptotic forms of fields are invariant is by looking at the invariance of metric. Under diffeomorphisms metric transforms in the following way
$$
\d_0g_{\m\n}=-\xi^\r\pd_\r g_{\m\n}-\pd_\m\xi^\r g_{\r\n}-\pd_\n\xi^\r g_{\m\r}\,.
$$
Let us note that metric does not transform under local Lorentz rotations.

Invariance of the vielbein gives a preliminary result for the sub-algebra of local Lorentz transformations that respect the asymptotic conditions, while invariance of spin connection may and will give further restrictions.

Transformation  law of the triad fields under diffeomorphisms and local Lorentz rotations is of the form
$$
\d_0 e^i_{\ \m}=-\varepsilon^{ijk}e_{j\m}\th_k-\left(\pd_\m\xi^\r\right) e^i_{\ \r}-\xi^\r\pd_\r e^i_{\ \m}\,.
$$

The  local symmetries that preserve the asymptotic form of vielbein are parametrized by
\bsubeq\lab{3.9}
\bea
&&\xi^t=T(t)+\cO_2\,,\quad
\xi^r=rU(\vphi)+\cO_1\,,\quad\xi^\varphi=S(\vphi)+\cO_3\,,\\
&&\th^0=-\frac{4r^2}{\ell^2}\pd_r\xi^t+\cO_3\,,\quad
\th^1= U(\vphi)+\pd_t T(t)+\cO_2\,,\quad\th^2=\th^0+\mathcal{O}_3\,.
\eea
\esubeq
By inspecting the invariance conditions for the asymptotic form of the spin connection
$$
\d_0 \omega^i_{\ \m}=-\nab_\m\th^i-\left(\pd_\m\xi^\r\right)\omega^i_{\ \r}-\xi^\r\pd_\r\omega^i_{\ \m}\,,
$$
we conclude that only the invariance of $\omega^1_{\ t}$ gives further restriction $\pd_t^2 T=0$ with the  simple solution
\be
T=At+B\,,
\ee
where $A$ and $B$ are constants.

\subsection{Asymptotic symmetry}

In order to find the interpretation of the asymptotic parameters, we calculate the commutator
algebra of the corresponding gauge transformations. First, we observe that commutator
algebra of the local Poincar\'e transformations  is closed: $[\d_0(1), \d_0(2)] = \d_0(3)$,
where $\d_0(1) := \d_0(\xi_1^\m,\th_1^i)$ etc, while the composition rule is given by:
\bea
&&\xi^\m_3=\xi_1\cdot\pd\xi^\m_2-\xi_2\cdot\xi_1^\m\,,\nn\\
&&\th_3^i=\ve^i{}_{mn}\th_1^m\th_2^m+\xi_1\cdot\pd\th_1^i-\xi_2\cdot\pd\th_2^i\,.\nn
\eea
After substituting the asymptotic parameters \eq{3.9} and comparing the lowest order terms we get:
\bea\lab{3.11}
&&U_3=S_1\pd_\vphi U_2-S_2\pd_\vphi U_1\,,\nn\\
&&S_3=S_1\pd_\vphi S_2-S_1\pd_\vphi S_1\,,
\eea
while $A_3=0$ and $B_3=A_1B_2-A_2B_1$. Motivated by this result in what follows we shall assume that $A=0$.

The pure gauge transformations are defined as the transformations generated by the higher order terms in \eq{3.9}
and they are irrelevant in the canonical analysis of the asymptotic
structure of spacetime \cite{x16}. The asymptotic
symmetry group is defined as the factor group of gauge transformations generated by \eq{3.9},
with respect to the residual gauge transformations.

Now, by introducing the Fourier expansion of the parameters and the notation:
\bea
&&j_n=\d_0(U=e^{in\vphi},S=0)\,,\nn\\
&&\ell_n=\d_0(U=0,S=e^{in\vphi})\,,\nn
\eea
we get that the asymptotic algebra takes the form of the semi-direct sum of $u(1)$ Kac-Moody and Virasoro algebra:
\bea\lab{3.12}
&&i[j_n,j_m]=0\,,\nn\\
&&i[j_m,\ell_n]=mj_{m+n}\,,\nn\\
&&i[\ell_m,\ell_n]=(m-n)\ell_{m+n}\,.
\eea
Central charges are absent, but their appearance can be
investigated  within canonical formalism, as we shall see in the next subsection.

\subsection{Charges}
The previous derivation is purely kinematical and is valid in any theory of gravity that has AdS orbifold with torsion as a solution. To conclude whether these  symmetries are  true ones or pure gauge we need to chose specific theory and calculate the charges that generate them. We decided to use canonical approach to asymptotic symmetries \cite{x17}.

\prg{Improved canonical generator.} Canonical charges are obtained by requiring that generator of gauge symmetry has well-defined functional derivatives for the given asymptotic behavior of fields. Generally, this leads to adding of a surface term $\G$ to the generator of symmetry $G$
\be
\tG=G+\G\,.
\ee
In this way we obtain the improved generator $\tilde{G}$ \cite{x17}.

After a shorter calculation we obtain that surface term needed that has to be added to a time translations generator is
\bea
\G[\xi^t]&=&2\int d\vphi \xi^t[e^2_{\ t}(a\omega^2_{\ \varphi}+\a_4e^2_{\ \varphi})+\omega^2_{\ t}(ae^2_{\ \varphi}+\a_3\omega^2_{\ \varphi})\nn\\
&-&e^0_t(a\omega^0_{\ \varphi}+\a_4e^0_{\ \varphi})-\omega^0_{\ t}(ae^0_{\ \varphi}+\a_3\omega^0_{\ \varphi})]\,.
\eea
In the same manner we obtain that surface term for symmetry generated by $\xi^\varphi$ is
\be
\G[\xi^\vphi]=-\int_{0}^{2\pi}d\varphi S(\varphi)\left(2ae^i_{\ \varphi}\omega_{i \varphi}+\a_4e^i_{\ \varphi}e_{i \varphi}+\a_3\omega^i_{\ \varphi}\omega_{i \varphi}\right)\,.
\ee
For $\xi^r$ we obtain the following surface term
\be
\G[\xi^r]=\int_{0}^{2\pi}d\varphi U(\varphi)\left[e^1_{\ \varphi}\left(\ell \a_4+a\left(\frac{p\ell}{2}-3\right)\right)+\omega^1_{\ \varphi}\left(a\ell +\a_3\left(\frac{p\ell}{2}-3\right)\right)\right]\,.\lab{3.16}
\ee
The generator of local Lorentz rotations is regular.

\prg{Canonical algebra.} Let us now find the Poisson bracket (PB) algebra of the the improved canonical generators. First, we introduce the notation
$\tG(1):= ˜\tG[U_1, S_1]$, $\tG(2):= ˜\tG[U_2, S_2]$, and we use the main
theorem of \cite{x18} to conclude that the PB $\{˜\tG(2),\tG(1)\}$ of two differentiable generators is
also a differentiable generator. This implies:
\be
\left\{\tG(2),\tG(1)\right\}=\tG(3)+C_{(3)}\,,
\ee
where the parameters of $\tG(3)$ are defined by the composition rule \eq{3.11}, while $C_{(3)}$ is an
unknown field-independent functional, $C_{(3)} := C_{(3)}[U_1, S_1; U_2, S_2]$, the central term of the
canonical algebra. The form of $C_{(3)}$ can be found using the relation
\be
\d_0(1)\G(2)\approx \G(3) + C_{(3)}\,.
\ee
We get
\be
C_{(3)}=\left[a\ell+\a_3\left(\frac{p\ell}2-1\right)\right]\int_0^{2\pi} d\vphi U_2\pd_\vphi U_1
\ee
After expressing the canonical generator
in terms of Fourier modes:
\be\lab{3.20}
J_n:=\tG[U=e^{in\vphi},S=0]\,,\qquad L_n:=\tG[U=0,S=e^{in\vphi}]\,,
\ee
the canonical algebra takes the form:
\bea\lab{3.21}
&&i\left\{J_m,J_n\right\}=\frac{\k}{12} m\d_{m+n,0}\,,\nn\\
&&i\left\{J_m,L_n\right\}=mJ_{m+n}\,,\nn\\
&&i\left\{L_m,L_n\right\}=(m-n)L_{m+n}\,,
\eea
where
\be
\kappa=24 \pi\left[a\ell+\a_3\left(\frac{p\ell}{2}-1\right)\right]\,.
\ee
Thus, the canonical realization of the asymptotic symmetry is given as the semi-direct sum
of $u(1)$ Kac-Moody algebra with central charge  and the Virasoro algebra
without central extension.

The values of the lowest order Kac-Moody and Virasoro generators take the form
\bea
&&L_0^{on-shell}=\frac{4\pi r^2_0}{\ell^2}\left[a\ell+\a_3\left(\frac{p\ell}{2} +1\right)\right]\,,\nn\\
&&J_0^{on-shell}=0\,.
\eea

\subsection{Alternative coordinates}
We can, alternatively, introduce the  coordinates $r=\dis\frac{\rho^2}{\ell}$
in which the metric of the orbifold reads
\be
ds^2=\frac{2r_0\rho^2}{\ell^2}dtd\vphi-\frac{\ell^2}{\rho^2}d\rho^2-r^2_0d\varphi^2\,.
\ee
where we further rescaled the time coordinate $t\ra\frac t2$.
Vielbeins are chosen in the following form
\bea
&&e^0=\frac{\rho^2}{\ell^2}dt\,,\qquad e^1=\frac{\ell}{\r}d\r\,,\qquad e^2=\frac{\r^2}{\ell^2}dt-r_0d\vphi\,.
\eea
In these coordinates metric takes the same form as that of near-horizon of rotating Oliva Tempo Troncoso (OTT) black hole \cite{x19}, so we can introduce the same asymptotic conditions on the metric
\be\lab{3.6}
g_{\m\n}\sim\left(
\ba{ccc}
\cO_{-1}&\cO_{3}&\cO_{-2}\\
\cO_{3}&\dis
-\frac{\ell^2}{\r^2}+\cO_4&\cO_1\\
\cO_{-2}&\cO_1&\cO_0
\ea
\right)\,.
\ee
The asymptotic behavior of the triad  fields is also the same as for the rotating OTT
\bea\lab{3.7}
e^i{_\m}\sim\left(
\ba{ccc}
\dis\frac{\r^2}{\ell^2}+\cO_{1}&\cO_5&\cO_0\\
\cO_1&\dis\frac\ell \r+\cO_3&\cO_0\\
\dis\frac{\r^2}{\ell^2}+\cO_1&\cO_5&\cO_0
\ea
\right)
\eea
Asymptotic form of the spin connection is given by
\bea\lab{3.8}
\omega^i{_\m}\sim\left(
\ba{ccc}
-\dis\frac{\r^2}{\ell^3}\left(1-\frac{p\ell}2\right)+\cO_1&\cO_2&\cO_0\\
\cO_1&-\dis\frac{1}{\r}\left(1-\frac{p\ell}2\right)+\cO_2&\cO_0\\
-\dis\frac{\r^2}{\ell^3}\left(1-\frac{p\ell}2\right)+\cO_{1}&\cO_2&\cO_0
\ea
\right)
\eea
and differs from the rotating OTT in the falloff of the $\omega^1_{\ t}$ component and in the presence of  the
torsional part parametrized by $p$. The asymptotic parameters for these falloff conditions are given by

\bsubeq
\bea
&&\xi^t=At+B+\cO_3\,,\qquad\xi^\r=\r V(\vphi)+\cO_1\,,\qquad\xi^\vphi=S(\vphi)+\cO_4\,,\\
&&\theta^0=\cO_2\,,\qquad \theta^1=2V(\vphi)+\partial_t\xi^t+\cO_4\,,\nn\\
&&\theta^2=\th^0-\frac{\ell^2}{\r^2}\pd_t\xi^\r+\cO_5\,,
\eea
\esubeq
where $A$ and $B$ are constants. By using the same arguments as in the previous subsection
we take $A=0$.

The asymptotic algebra, obtained  after introducing the Fourier expansion of the parameters and the notation
\bea
&&k_n=\d_0(V=e^{in\vphi},S=0)\,,\nn\\
&&\ell_n=\d_0(V=0,S=e^{in\vphi})\,,\nn
\eea
 takes the form of the semi-direct sum of $u(1)$ Kac-Moody and Virasoro algebra
\bea
&&i[k_n,k_m]=0\,,\nn\\
&&i[k
_m,\ell_n]=mk_{m+n}\,,\nn\\
&&i[\ell_m,\ell_n]=(m-n)\ell_{m+n}\,.
\eea
It is isomorphic with algebra \eq{3.12}.

The  central charges  can be obtained in the similar manner by employing canonical formalism.
The surface terms for generators corresponding to $\xi^t$ and $\xi^\varphi$ are the same as for the previous asymptotic conditions, while the surface term for the generator $G[\xi^r]$ takes form
\be
\G[\xi^\r]=\int_{0}^{2\pi}d\varphi V(\varphi)[e^1_{\ \varphi}(2\ell \a_4+a(p\ell-6))+\omega^1_{\ \varphi}(2a\ell +\a_3(p\ell-6))]\,.\lab{3.31}
\ee
After introducing the Fourier modes of the improved canonical generators
\be
K_n:=\tG[V=e^{in\vphi},S=0]\,,\qquad L_n:=\tG[V=0,S=e^{in\vphi}]\,,\lab{3.32}
\ee
we get that the canonical algebra takes the form:
\bea\lab{3.32}
&&i\left\{K_m,K_n\right\}=\frac{\k'}{12} m\d_{m+n,0}\,,\nn\\
&&i\left\{K_m,L_n\right\}=mK_{m+n}\,,\nn\\
&&i\left\{L_m,L_n\right\}=(m-n)L_{m+n}\,,
\eea
where
\be
\kappa'=96 \pi\left[a\ell+\a_3\left(\frac{p\ell}{2}-1\right)\right]\equiv 4\k\,.
\ee
The values of the zero mode generators on the orbifold background are given by
\bea
&&L_0^{on-shell}=\frac{4\pi r^2_0}{\ell^2}\left[a\ell+\a_3\left(\frac{p\ell}{2} +1\right)\right]\,,\nn\\
&&J_0^{on-shell}=0\,.
\eea

Therefore, the canonical algebra \eq{3.32} is {\it isomorphic} to the canonical algebra
\eq{3.21}. The simple mapping that relates the  generators of the two algebras is given by
\bea
L_n\leftrightarrow L_n\,,\qquad K_n\leftrightarrow 2J_n\,.\lab{3.35}
\eea
The mapping \eq{3.35} can be easily  derived from the definitions of the Fourier modes of the improved generators
$K_n$ and $J_n$, given by equations \eq{3.32} and \eq{3.20}, and form of the corresponding surface terms
$\G(\xi^\r)$ and $\G(\xi^r)$, equations \eq{3.31} and \eq{3.16}.

\section{Concluding remarks}

We analyzed near-horizon geometry with non-trivial torsion with the aim to understand its influence  on applicability of Kerr/CFT. For simplicity, we undertook the investigation of simplest case of extremal BTZ black hole with torsion. After deriving the corresponding near-horizon geometry which is the generalization of AdS self-dual orbifold -- the near-horizon of BTZ black hole. After introducing the suitable asymptotic conditions we derived the algebra of asymptotic symmetries, which consists of the Killing vectors of the orbifold and direct sum of chiral Virasoro and $u(1)$ Kac-Moody algebra, known also as warped CFT symmetry. Virasoro algebra is not centrally extended, while Kac-Moody algebra possesses non-zero central extension $\kappa$.

It is worth noting that investigation of asymptotic structure of BTZ black hole with torsion was done in \cite{x11}  with the result that asymptotic symmetry is Virasoro algebra with central charges $c$ and $\bar{c}$. Central extension of Kac-Moody algebra, that we obtained, is proportional to $c$, and on-shell value of Virasoro zero mode generator is proportional to $\bar{c}$.

There is the procedure, known as Sugawara-Sommerfeld construction \cite{x20}, for constructing Virasoro algebra from bilinear combinations of Kac-Moody generators.  In this way it is possible to obtain Virasoro algebra which allows the application of the Cardy formula. The drawback of this approach is that one of the central charges is proportional to an arbitrary constant which is fixed by demand that Cardy formula correctly reproduces black hole entropy.

\section*{Acknowledgements}
This work was partially supported by the Serbian Science foundation, Serbia, grant 171031.

\appendix
\section{Proof that surface term $\G[\xi^t]$ is finite}
\setcounter{equation}{0}

Since, the surface term takes  the same form  for both choices of asymptotic conditions it is enough to show that it is finite for one of them.

In the case of radial coordinate $\rho$  the following relations between Riemannian part of connection and vielbein hold
\bsubeq
\bea
&&\frac{e^0_{\ \vphi}}{\ell}-\frac{e^2_{\ \vphi}}{\ell}+\tilde{\omega}^{2}_{\ \vphi} -\tilde{\omega}^{0}_{\ \vphi}=\cO_2\,,\lab{A.1a}\\
&&\tilde{\omega}^{2}_{\ t}=-\frac{e^2_{\ t}}{\ell}+\cO_1=-\frac{\rho^2}{\ell^3}+\mathcal{O}_1\,,\lab{A.1b}\\
&&\tilde{\omega}^{0}_{\ t}=-\frac{e^0_{\ t}}{\ell}+\cO_1=-\frac{\rho^2}{\ell^3}+\mathcal{O}_1\,.\lab{A.1c}
\eea
\esubeq
The first identity represents a consequence of  $T^i=0$, while the other two are obvious from the asymptotic behavior of fields.

If we expand the Lorentz spin connection as $\omega^i=\tilde{\omega}^i+\frac{p}{2}e^i$, and use the relation between Lagrangian parameters $\a_4=\frac{\a_3}{\ell^2}-ap-\frac{\a_3p^2}{4}$ as well as  equalities \eq{A.1b} and \eq{A.1c}, after a shorter calculation, we derive
\be
\G[\xi^t]=\int d\vphi \frac{\rho^2}{\ell^2}\left(2a+\a_3p-\frac{2\a_3}{\ell}\right)\left(\frac{e^0_{\ \vphi}}{\ell}-\frac{e^2_{\ \vphi}}{\ell}+\tilde{\omega}^{2}_{\ \vphi} -\tilde{\omega}^{0}_{\ \vphi}\right)+\cO_1\,.
\ee
Now, from the first relation \eq{A.1a} it follows that  $\G[\xi^t]=\mathcal{O}_0$, i.e. it is finite.

\end{document}